\documentclass[prl,showpacs,showkeys,amsmath,amssymb,twocolumn]{revtex4}

\usepackage{dcolumn}
\usepackage{bm}
\input xy
\usepackage[all]{xy}
\xyoption{2cell} \UseAllTwocells

\begin{document}



\title{Dirac Quantization and Fractional Magnetoelectric Effect
\\on Interacting Topological Insulators}
\author{K.-S. Park}
\email{kpark@postech.ac.kr}
\author{H. Han}
\affiliation{Department of Electrical Engineering, Pohang University
of Science and Technology, San 31, Hyojadong, Namgu, Pohang 790-784,
Korea}

\date{\today}

\begin{abstract}
We use Dirac quantization of flux to study fractional charges and
axion angles $ \theta $ in interacting topological insulators with
gapless surface modes protected by time-reversal symmetry. In
interacting topological insulators, there are two types of
fractional axion angle due to conventional odd and nontrivial even
flux quantization at the boundary. On even flux quantization in a
gapped time reversal invariant system, we show that there is a
halved quarter fractional quantum Hall effect on the surface with
Hall conductance of $ \frac{p}{4q} \frac{e^{2}}{2h} $ with $ p, q $
odd integers. The gapless surface modes can be characterized by a
nontrivial $ \mathrm{Z}_{2} $ anomaly emerged from the even flux
quantization. It is suggested that the electron can be regarded as a
bound state of fractionally charged quarks confined by a nonabelian
color gauge field on the Dirac quantization of complex spinor
fields.
\end{abstract}

\pacs{73.43.-f, 75.80.+q, 71.27.+a,11.15.-q}
\keywords{fractional topological insulator, surface Hall
conductance, fractional charge, axion angle, Dirac quantization,
topological field theory.}

\maketitle

Recently there has been paid a great attention to topologically
nontrivial states of quantum matter on time-invariant topological
insulators (TI) \cite{moore1,kane1,qz}. As a theory of 3D TI
theories \cite{kane2,moore2,roy,bernevig}, topological field
theories(TFTs) have been developed in the low energy limit on the
various dimensions of TIs \cite{qi1}. In the formulation of a
noninteracting TFT on (3 + 1)D, special quantum effects can be
induced from the interior of 3D TIs through the couplings between
electric and magnetic fields on the Dirac quantization condition of
charges and fluxes, similar to the coupling of an axion particle to
ordinary electric and magnetic fields. The theory of TIs is
described by the effective action
\begin{eqnarray}
S_{\theta}(\mathrm{E},\mathrm{B}) = K_{\theta} \frac{e^{2}}{2 \pi}
\int_{M_{4}} d^{3}x dt \mathrm{E} \cdot \mathrm{B}
\end{eqnarray}
where $ K_{\theta} = \frac{\theta}{2 \pi}$, $ \mathrm{E} $ and $
\mathrm{B} $ stand for the electromagnetic fields
\cite{wilczek2,qi1,essin,qi2}. On the condition of normalization all
physical quantities become preserved under shifts of $ \theta $ by
multiples of $ 2 \pi $. Because $ \mathrm{E} \cdot \mathrm{B} $ gets
odd under the T operation, there are 0 or $ \pi $ mod $ 2 \pi $ in
the values allowed by T reversal symmetry. Furthermore, the
time(T)-invariant fractional topological states have been suggested
in (4 + 1)D \cite{zhang}. Fractional states can, in general, be
appeared in terms of strong interactions in (3 + 1)D. Hence it is
very important to construct a general theory of fractional TIs on a
(3 + 1)D spin manifold in the presence of strong interactions.

In this brief report, we exploit Dirac quantization to investigate
fractional charges and axion angles $ \theta $ under the
construction of general theory for interacting TIs with gapless
surface modes protected by time-reversal symmetry. For interacting
TIs, we report three crucial results. First, there exist two kinds
of fractional axion angle due to the conventional odd and nontrivial
even flux quantization at the boundary. Secondly, when making the
even flux quantization required in a gapped time reversal invariant
system, we show that there is a halved quarter fractional quantum
Hall effect(FQHEs) on the surface with Hall conductance of $
\mathrm{\sigma}_{H}^{s} = \frac{p}{4q} \frac{e^{2}}{2h} $ with $ p,
q $ odd integers. Thus it is proposed that the Dirac quantization of
even flux leads to the fractional bulk topological quantum number
for non-integer, rational multiples such that $ K_{\theta} =
\frac{1}{2} \frac{p}{4q} $ with $ p,q $ odd integers on the (3 + 1)D
spin manifold of fractional TIs. And finally we claim that the
system can have degenerate ground states on a closed topologically
nontrivial space- and time-reversal protected gapless surface states
which are characterized by $ \mathrm{Z}_{2} $ anomaly of $
(-1)^{\omega(\Sigma)} $ with $ \omega(\Sigma) $ mod 2 of a 2-cocycle
$ \Sigma $, i.e., 2-cycle of a closed form, caused by even flux
boundary. In a time reversal (TR) symmetric TI of quantum magnets,
there can be two topological objects of fractionalized charge $
\frac{1}{2} $ emerged from an interacting TI with a band gap because
of the even flux quantization which is twice a 1-cycle flux
quantization. Very recently, Maciejko and his companies have
suggested the possibility of TR invariant fractional topological
insulators for fermions in strong coupled $ SU(N) $ gauge theory
\cite{maciejko}. In the large $ N $ limit, this theoretical
construction can have a serious problem due to a TR symmetry broken
spontaneously because $ SU(N) $ gauge theory can not preserve time
reversal symmetry \cite{witten0}. But within our theoretical
framework of interacting TIs, fractional topological insulators can
be realized in the absence of TR symmetry breaking caused by strong
interacting $ SU(N) $ gauge theory. The current theory can only give
rise to fractional $ \frac{\theta}{\pi} $ in a gapped time reversal
invariant system of bosons or fermions provided that the system also
takes deconfined fractional excitations and associated degenerate
ground state on topologically nontrivial spaces. We show the above
findings theoretically as followings.

In order to investigate the fractional TI on the basis of a more
systematical approach given by Maciejko and his companies, we take
into account the projective construction of FQH states for a
composite electron \cite{maciejko,jain,wen,barkeshli} on the surface
of an emerged spin manifold in (3 + 1)D. The electron is decomposed
into $ N_{f}^{c} $ different flavors of fractionally charged and
fermionic partons with $ N_{f}^{c} $ partons of each flavors $ f =
1, \cdots, N_{f} $. This decomposition should obey two fundamental
constraints:
\begin{eqnarray}
 N_{1}^{c} + \cdots + N_{N_{f}}^{c}  = \mathrm{odd}, \quad
 N_{1}^{c} q_{1} + \cdots + N_{N_{f}}^{c}q_{N_{f}} = e.
\end{eqnarray}
The first constraint of Eq. (2) means that the total number of
partons per electron has to be odd because the electron preserves
the fermion statistics. The other constraint is that the total
charge of the partons has to sum up to the electron charge $ e $
when $ q_{f} < e $ becomes the fractional charge for partons of
flavor $ f $.

Provided that the partons get recombined together to represent the
physical electrons, we can construct an interacting many-body
wavefunction as a new topological state of electrons emerged in (3 +
1)D. The total electron wavefunction is expressed by a product of
parton ground state wavefunctions \cite{jain}
\begin{eqnarray}
\prod_{f=1}^{N_{f}} \Psi_{N_{f}^{c}}(
\{\mathrm{r}_{i},\mathrm{s}_{i}\} ) =\prod_{f=1}^{N_{f}} \lbrack
\Psi_{f}( \{ \mathrm{r}_{i},\mathrm{s}_{i} \} ) \rbrack^{N_{f}^{c}}.
\end{eqnarray}
Here $ \Psi_{f}( \{\mathrm{r}_{i},\mathrm{s}_{i}\} ) $ stands for
the parton ground state wavefunction given by a Slater determinant
which describes the ground state of a noninteracting TI Hamiltonian,
and $ \{\mathrm{r}_{i},\mathrm{s}_{i}\}, i = 1, \cdots, N $, the
position and spin coordinates of the partons.

To be more specific, let us consider an effective field theory of
fractional TI on a diamond lattice of the $ SU(N) $ electrons. The
Hamiltonian is given by
\begin{eqnarray}
H = \sum_{ab} \{ C^{\dagger}_{a \alpha} h^{\alpha \beta}_{ab}
e^{ieA_{ab}} C_{b \beta} + H. C. \} + H_{int}(C^{\dagger},C),
\end{eqnarray}
where $ a, b $ indicate site indices, $ \alpha, \beta $ stand for
internal degrees of freedom, $ h_{ab} $ denotes the Hamiltonian
matrix, $ A_{ab} = \int_{\mathrm{r}_{a}}^{\mathrm{r}_{b}}
d\mathrm{r} \cdot \mathrm{A} $ with $ \mathrm{A} $, the $ U(1) $
electromagnetic vector potential, and $ H_{int} $ means an
interaction Hamiltonian between electrons. $ C_{a \alpha} $ is the
electron operator decomposed as
\begin{eqnarray}
C_{a \alpha} = \prod_{f=1}^{N_{f}} \psi^{f}_{1
\alpha}(\mathrm{r}_{a}) \cdots \psi^{f}_{ N_{f}^{c} \alpha
}(\mathrm{r}_{a}) \end{eqnarray} with obeying constraint rules Eq.
(2). Here $ \psi^{f}_{j \alpha}(\mathrm{r}_{a}), j = 1, \cdots,
N_{f}^{c} $ are quark operators with $ N_{f}^{c} $ partons of each
flavor $ f $. The projection onto the physical Hilbert space can be
realized by including $ SU(N_{1}^{c} $ gauge transformation.

Now we take into account the Dirac quantization \cite{dirac} of
fermions generated by complex spinor fields through a two-cycle $
\Sigma $ in the sense of antisymmetric $ N^{c}_{1} $ partons, i.e.,
the odd-number constraint of $ N^{c}_{1} $, of a composite electron
on $ M_{4} $. The extension of Dirac quantization to any 2-cycle was
described by O. Alvarez \cite{alvarez,am}. Let us study the Dirac
quantization by following the Alvarez's extension. Under the Dirac
quantization, the antisymmetric parton wavefunctions can be
represented by a nonabelian color gauge field $ SU(N_{1}^{c}) $ with
an interacting constant $ g $ on $ M_{4} $ which is covered by a
finite number of neighborhoods $ U_{i}, i=1,\cdots, N $. In each
neighborhood, more structures should be considered on the
representation of internal symmetries. In addition to the $ U(1) $
connection or gauge potential, there must be an oriented frame of
vierbein $ V_{i} $ and complex spinor fields of antisymmetric
partons $ \{ \Psi_{1i} \}^{N_{1}^{c}} $ with $ N_{1}^{c} $ only odd.
These symmetries cannot be independent of choices made in the
neighborhood $ U_{i}$. As choices of degrees of freedom, there can
be local $ U(1) \times SU(N_{1}^{c}) $ gauge transformations $
(\mathit{\chi},\mathit{\lambda}) $
\begin{eqnarray}
\{ \Psi_{1i} \}^{N_{1}^{c}} \rightarrow \{ e^{iq_{1} \mathit{\chi}
+ig \mathit{\lambda} } \Psi_{1i} \}^{N_{1}^{c}}, \nonumber\\
 A_{i} \rightarrow A_{i} + d \mathit{\chi}, \quad a_{i} \rightarrow a_{i} +
d \mathit{\lambda},
\end{eqnarray}
and $ SO(4) $ local transformations \cite{am}
\begin{eqnarray}
V_{i} \rightarrow \mathrm{R} V_{i}, \quad \{ \Psi_{1i}
\}^{N_{1}^{c}} \rightarrow \{ \mathrm{S}( \mathrm{R} ) \Psi_{1i}
\}^{N_{1}^{c}},
\end{eqnarray}
where $ \mathrm{R}  \in SO(4) $. It is easy to see that there can be
a sign ambiguity from the lift of $ \mathrm{R} \rightarrow \pm
\mathrm{S}( \mathrm{R} ) $ since the quotient of the spin group $
Spin(4)$ by $ \mathrm{Z}_{2} $ is isomorphic to $ SO(4) $. For a
double overlap on two contiguous neighborhoods, $ U_{i} \cap U_{j}
\ne 0 $, one should take transition functions associated with
transformation groups
\begin{eqnarray}
A_{i} \rightarrow A_{j} + d \mathit{\chi}_{ij}, \quad a_{i}
\rightarrow a_{j} + d \mathit{\lambda}_{ij}, \quad
V_{i} \rightarrow \mathrm{R}_{ij} V_{j}, \nonumber \\
\{ \Psi_{1i} \}^{N_{1}^{c}} \rightarrow \{ \mathrm{S}(
\mathrm{R}_{ij} )e^{iq_{1} \mathit{\chi}_{ij}} e^{ig
\mathit{\lambda}_{ij}} \Psi_{1j} \}^{N_{1}^{c}}.
\end{eqnarray}
In a triple overlap region, $ U_{i} \cap U_{j} \cap U_{k} \ne 0$,
which is supposed to be contractible, we can have consistency
conditions
\begin{eqnarray}
 \mathrm{R}_{ij}\mathrm{R}_{jk}\mathrm{R}_{ki} = I,
 S(ijk) \equiv \mathrm{S}( \mathrm{R}_{ij} ) \mathrm{S}(
\mathrm{R}_{jk} ) \mathrm{S}( \mathrm{R}_{ki} ) = \pm I. \quad
 \end{eqnarray}
It is noted that the above equations have identity elements of $
SO(4) $ and $ Spin(4)$. Consequently, under $ U(1) \times
SU(N_{1}^{c}) $ gauge transformations and in the spinor
representations of antisymmetric $ N^{c}_{1} $ partons with $
N^{c}_{1} $ odd, we can obtain
\begin{eqnarray}
\{ \Psi_{1i} \}^{N_{1}^{c}} &=& \{ e^{iq_{1} ( \mathit{\chi}_{ij} +
\mathit{\chi}_{jk} + \mathit{\chi}_{ki} )} e^{ig
(\mathit{\lambda}_{ij} + \mathit{\lambda}_{jk} +
\mathit{\lambda}_{ki} )} \nonumber\\
&& \mathrm{S}( \mathrm{R}_{ij} )
\mathrm{S}( \mathrm{R}_{jk} ) \mathrm{S}( \mathrm{R}_{ki} )
\Psi_{1i} \}^{N_{1}^{c}}.
\end{eqnarray}
Up to a sign, a crucial point is to take a lift from $ SO(4) $ to $
Spin(4) $ in the right hand side of Eq. (10). We cannot determine
the sign when it is dependent on the choices made in the
transformation groups of Eq. (10). In the sense of different
overlaps, the signs of Eq. (10) cannot be totally independent. Thus
the spinor consistency condition enables us to obtain
\begin{eqnarray}
e^{iq_{1} C_{ijk}} e^{ig D_{ijk}} = S(ijk),
\end{eqnarray}
in the model of total $ U(1) \times SU(N_{1}^{c}) $ gauge fields.
Here $ C_{ijk} $ and $ D_{ijk} $ satisfy the self-consistency
relations
\begin{eqnarray}
C_{ijk} \equiv \mathit{\chi}_{ij}(\mathrm{r}) +
\mathit{\chi}_{jk}(\mathrm{r}) + \mathit{\chi}_{ki}(\mathrm{r}) \in
\frac{\mathrm{Z}}{q_{1}N_{1}^{c}}, \nonumber\\
D_{ijk} \equiv \mathit{\lambda}_{ij} + \mathit{\lambda}_{jk} +
\mathit{\lambda}_{ki} \in  \frac{\mathrm{Z}}{g N_{1}^{c}}.
\end{eqnarray}
Therefore, by $ U(1) \times SU(N_{1}^{c}) $ gauge theory through a
two-cycle $ \Sigma $ which is a 2D-manifold without boundary, the
Dirac quantization of flux gives rise to
\begin{eqnarray}
\mathrm{exp}( 2\pi i \int_{\Sigma} ( q_{1} F + g G ) ) =
(-1)^{\omega(\Sigma)},
\end{eqnarray}
where $ G $ is the $ SU(N_{1}^{c}) $ field strength, with $
N_{1}^{c} $ odd. Here the sign is determined by the finite product
over triple overlaps
\begin{eqnarray}
(-1)^{\omega(\Sigma)} = \prod_{U_{i} \cap U_{j} \cap U_{k} \cap
\Sigma \ne 0} S(ijk).
\end{eqnarray}

Finally let us account for the flux through $ \Sigma_{i} $ with
nontrivial boundary $ \partial \Sigma_{i}, \forall i = 1, 2, \dots,
N $. Then in the representation of complex spinor fields for
fermions generated by antisymmetric partons, the spinor field
consistency in Eq. (11) leads to the boundary Dirac quantization
condition
\begin{eqnarray}
&&\mathrm{exp}( 2\pi i \int_{\Sigma_{i}} ( q_{1} F + g G ) ) =
\mathrm{exp}( 2\pi i \oint_{\partial \Sigma_{i}} ( q_{1} A + g a ) ) \nonumber\\
&& \times \prod_{U_{i} \cap U_{j} \cap U_{k} \cap \Sigma \ne 0}
S(ijk), \quad \forall i = 1, 2, \dots, N.
\end{eqnarray}
The problem in question is that the two factors of Eq. (15) cannot
be independent of the choice of neighborhoods although the
neighborhoods can be independently chosen in the right hand side.
Let us consider Dirac flux quantization at the conventional odd
boundary of a 4D manifold. In the viewpoint of conventional flux
quantization at the odd boundary, i.e., $ \partial \Sigma_{i} =
\gamma_{i} $ which means a loop or a 1-cycle, it is argued that
adding a neighborhood to the interior of $ \Sigma $ cannot affect
the first factor as well as the second factor. Thus the boundary
Dirac quantization yields to
\begin{eqnarray}
&&\mathrm{exp}( 2\pi i \int_{\Sigma_{i}} ( q_{1} F + g G ) )
\nonumber\\
&=& \mathrm{exp}( 2\pi i \oint_{\gamma_{i}} ( \frac{e}{N_{1}^{c}} A
+ g a ) ), \forall i = 1, 2, \dots, N
\end{eqnarray}
Therefore we can find $ q_{1} = \frac{e}{N_{1}^{c}} $ due to a
conventional odd Dirac quantization of flux of antisymmetric $
N_{1}^{c} $ partons at the boundary, i.e., $
\partial \Sigma_{i} = \gamma_{i} $ on a 4D manifold.

Although any better way is not known for understanding Stokes'
theorem in the current context, the sign problem in Eq. (15) will,
however, be resolved if the boundary of $ \Sigma_{i} $ become even
such as $
\partial \Sigma_{i} = 2 \gamma_{i}, \forall i = 1, 2, \dots, N $.
Hence the even flux quantization leads to a form \cite{alvarez,am}
\begin{eqnarray}
&&\mathrm{exp}( 2\pi i \int_{\Sigma_{i}} ( q_{1} F + g G ) ) \nonumber\\
&=&(-1)^{\omega(\Sigma_{i})}\mathrm{exp}( 2\pi i \oint_{ \gamma_{i}}
2( q_{1} A + g a ) ), \forall i = 1, 2, \dots, N \qquad
\end{eqnarray}
If the two factors of Eq. (15) become independent of choices of
neighborhoods, they can be well defined since the second factor is
well-behaved due to the extra number 2 in the exponent of Eq. (17).
Therefore we can obtain $ q_{1} = \frac{e}{2 N_{1}^{c}} $ from the
even flux quantization at the boundary, i.e. $
\partial \Sigma = 2 \gamma $ on a 4D manifold with spin structures.
It follows that $ \omega(\Sigma) $ becomes well-defined mod 2 on the
$ \Sigma $ with even flux boundary condition. This remarkable result
shows that the system can have degenerate ground states on a closed
topologically nontrivial space-time-reversal-protected gapless
surface states which are characterized by $ (-1)^{\omega(\Sigma)} $
at the even flux boundary. There can be two topological objects of
fractionalized charge $ \frac{1}{2} $ emerged from the even flux
quantization.

So far we have described the Dirac quantization of flux with the
total $ U(1) \times SU(N_{1}^{c}) $ quark field strength $ q_{1} F +
g G $ on the complex spinor representations of antisymmetric parton
wavefunctions. The interactions yield the quarks to condensing at
low energies into a noninteracting T-invariant TI state with axion
angle $ \theta $. It has been assumed that $ M _{4} $ has spin
structures, and odd and even boundaries. In particular, we have to
account for the three main results of boundary flux quantization
into the topological term of the effective field theory. In the
topological term $ \frac{\theta}{2 \pi} \frac{e^{2}}{2 \pi}
\mathrm{E} \cdot \mathrm{B} = \frac{\theta e^{2}}{32 \pi^{2}}
\epsilon^{\mu \nu \rho \sigma} F_{\mu \nu}F_{\rho \sigma} $ for
noninteracting TIs, the $ U(1) $ electron field strength is replaced
by the $ U(1) \times SU(N_{1}^{c}) $ one. We can construct a
partition function
\begin{eqnarray}
\mathcal{Z} = C (-1)^{\omega(\Sigma)} \mathrm{exp}( i\int_{M_{4}}
d^{3}x dt \mathcal{L}_{eff} (F,G) ).
\end{eqnarray}
The effective Lagrangian is given by $ \mathcal{L}_{eff} (F,G) =
\mathcal{L}_{0} + \mathcal{L}_{top} $ where $ \mathcal{L}_{0} =
-\frac{1}{4}G_{\mu \nu}G^{\mu \nu} $ is the kinetic Yang-Mills (YM)
Lagrangian. The second topological term of $ \mathcal{L}_{eff} (F,G)
$ is expressed in terms of
\begin{eqnarray}
&&\mathcal{L}_{top} = \frac{\theta_{1}}{32 \pi^{2}} \epsilon^{\mu
\nu \rho \sigma} \mathrm{Tr}[( q_{1} F_{\mu \nu} + g G_{\mu \nu})(
q_{1} F_{\rho \sigma} + g G_{\rho \sigma} )]
\nonumber\\
&=& \partial_{\mu} \epsilon^{\mu \nu \rho \sigma} \{
\frac{\theta_{eff} e^{2}}{8 \pi^{2}}A_{\nu}\partial_{\rho}A_{\sigma}
+ \frac{\theta_{1} g^{2}}{8 \pi^{2}}
(a_{\nu}\partial_{\rho}a_{\sigma} + \frac{2}{3}
iga_{\nu}a_{\rho}a_{\sigma}) \}, \qquad
\end{eqnarray}
where $ \mathrm{Tr} $ stands for the trace in the $ N_{1}^{c} $
representation of $ SU(N_{1}^{c}) $, $ F_{\mu \nu} = \partial_{\mu}
A_{\nu} - \partial_{\nu} A_{\mu} $ and $ G_{\mu \nu} =\partial_{\mu}
a_{\nu} - \partial_{\nu} a_{\mu} + ig[a_{\mu},a_{\nu}] $ indicate $
U(1) $ and $ SU(N_{1}^{c}) $ field strengths, respectively, and $
\theta_{1} $ is an action angle for $ N_{f}$ =1. It is noted that
the crossed terms such as $ \mathrm{Tr}( F_{\mu \nu}f_{\rho \sigma})
$ become zero owing to the tracelessness of the $ SU(N_{1}^{c}) $
gauge field. Under the nonabelian gauge theory , there can possibly
be two phases such as confined phases with chiral symmetry breaking
and deconfined phases with gapless gauge modes for large flavor
cases. In this work, we however focus on the deconfined phases for
interacting TIs with topologically protected surface states, and the
YM kinetic term does not make a crucial contribution in the large
flavor cases of TIs.

The electromagnetic response allows us to have two classes of
effective axion angles obtained from conventional odd and nontrivial
even boundary Dirac quantization caused by Eq. (16) and Eq. (17),
respectively. From the first term of Eq. (19), they are given,
respectively, by
\begin{eqnarray}
\theta_{eff} &=& \frac{\theta_{1}}{N_{1}^{c}} = 0, \pm
\frac{\pi}{N_{1}^{c}},\pm \frac{3\pi}{N_{1}^{c}}, \pm
\frac{5\pi}{N_{1}^{c}}, \cdots ,  N_{1}^{c} \quad \mathrm{odd},
\quad
\nonumber\\
\theta_{eff} &=& \frac{\theta_{1}}{2N_{1}^{c}} = 0, \pm
\frac{\pi}{2N_{1}^{c}},\pm \frac{3\pi}{2N_{1}^{c}}, \pm
\frac{5\pi}{2N_{1}^{c}}, \cdots , N_{1}^{c} \quad \mathrm{odd}.
\qquad
\end{eqnarray}
According to the first and second term in Eq. (19), the T-invariance
enables us to quantize $ \theta_{eff} $ in integer multiples of $
\pi $ if the minimal electric charge becomes $ e $, and in the
current parton model of quarks, $ \theta_{1} $ is $
\frac{e}{N_{1}^{c}} $ \cite{maciejko} and $ \frac{e}{2N_{1}^{c}}$
due to odd and even flux quantization at the boundary, respectively.
But conventional odd flux quantization does not lead to a time
reversal protected topological insulator because $ SU(N_{1}^{c}) $
gauge theory has a serious TR symmetry breaking in the large $
N_{1}^{c} $ limit \cite{witten0}. On the other hand, owing to $
\mathrm{Z}_{2} $ anomaly of $ (-1)^{\omega(\Sigma)} $ in Eq. (18)
with $ \omega(\Sigma) $ mod 2 of a 2-cocycle $ \Sigma $ at the
boundary, it follows that even flux quantization can yield a
topological insulator of TR invariance. Consequently, $ \theta_{eff}
$ should have quantized values in odd integer multiples of $
\frac{\pi}{(2N_{1}^{c})^{2}} $ as a crucial result obtained in the
current theory.

Let us consider the effective field theory in the multiple flavor
values $ N_{f}^{c} \ge 1 $. Assume that quarks of flavor $ f $
produces a noninteracting TI with $ \theta_{f} = \pi $ $
\mathrm{mod} $  $ 2\pi $. Then after integrating them, the effective
theory has a gauge group of $ U(1) \times \prod_{f=1}^{N_{f}^{c}}
U(N_{f}^{c})/U_{e}(1) $ where $ U(1)_{e} $ denotes the overall $
U(1) $ gauge transformation of the electron operator. This gauge
group leads to the axion angle $ \theta_{eff} = \{
\sum_{f=1}^{N_{f}} \frac{N_{f}^{c}}{\theta_{f}} \}^{-1} $. If $
\frac{\theta_{f}}{\pi} $ becomes odd for each flavor, then there can
be $ p $ and $ q $ of odd integers such as $ \theta_{eff} = \pi
\frac{p}{4q} $. The partition function of the gauge fields allows us
to have important physical properties for the interacting TI. In
general, the surface of the interacting TI has an action domain wall
with the electromagnetic action angle which jumps from $
\theta_{eff} $ in the fractional TI to 0 in the vacuum. Hence by
even boundary Dirac quantization we can find a halved quarter FQHE
on the surface with Hall conductance
\begin{eqnarray}
\sigma_{H}^{s} = \frac{p}{4q} \frac{e^{2}}{2h}, \quad p, q &
\mathrm{odd}.
\end{eqnarray}

In conclusion and summary, we have exploited two kinds of Dirac
quantization of flux to investigate fractional charges and axion
angles on topological insulators. For an interacting TI, the
nontrivial even flux quantization at the boundary yields a halved
quarter fractional QHE with Hall conductance in Eq. (21) due to a
topological origin on a 4D manifold with spin structures. The system
can have degenerate ground states on a closed topologically
nontrivial space and time reversal protected gapless surface states
characterized by a factor $ (-1)^{\omega(\Sigma)} $ emerged from the
even flux quantization. Hence it is proposed that fractional TIs can
be realized within the current interacting TI theory which may
provide the nontrivial properties of the quantum state with
fractionalization and the emergent nonabelian gauge theory without
the microscopic models. These results can be obtained in the
strongly correlated systems of strong spin-orbit interaction
\cite{park1,swingle}.

\begin{acknowledgments}
This work was supported by the National Research Laboratory Program
(R0A-2005-001-10152-0), by Basic Science Research Program
(2009-0083512), and by Priority Research Centers Program
(2009-0094037) through the National Research Foundation of Korea
funded by the Ministry of Education, Science and Technology, and by
the Brain Korea 21 Project in 2010.
\end{acknowledgments}


\begin{thebibliography}{10}

\bibitem{moore1}
J. E. Moore, Nature {\bf 464}, 194 (2010); Nature {\bf 460}, 1090
(2008);

\bibitem{kane1}
M. Z. Hasan {\it et al. }, http://arXiv.org/abs/1002.3895v1.

\bibitem{qz}
X. L. Qi {\it et al. }, Phys. Today {\bf 63}, 33 (2010).

\bibitem{kane2}
C. L. Kane and E. J. Mele, Phys. Rev. Lett. {\bf 95}, 146802 (2005);
Phys. Rev. Lett. {\bf 98}, 106803 (2007).

\bibitem{moore2}
J. E. Moore {\it et al. }, Phys. Rev. B. {\bf 75}, 121306(R) (2007).

\bibitem{roy}
R. Roy, Phys. Rev. B. {\bf 79}, 195322 (2009).

\bibitem{bernevig}
B. A. Bernevig {\it et al. }, Science {\bf 314}, 1757-1761 (2006).

\bibitem{qi1}
X.-L. Qi {\it et al. }, Phys. Rev. B. {\bf 78}, 195424 (2008).

\bibitem{wilczek2}
F. Wilczek, Phys. Rev. Lett. {\bf 58}, 1799 (1987).

\bibitem{essin}
A. M. Essin {\it et al. }, Phys. Rev. Lett. {\bf 102}, 146805
(2009).

\bibitem{qi2}
X.-L. Qi {\it et al. }, Science {\bf 323}, 1184 (2009).

\bibitem{zhang}
S. C. Zhang and J. P. Hu, Science {\bf 294}, 823 (2001).

\bibitem{maciejko}
J. Maciejko {\it et al. }, http://arXiv.org/abs/1004.3628v1.

\bibitem{witten0}
E. Witten, Phys. Rev. Lett. {\bf 81}, 2862 (1998).

\bibitem{jain}
J. K. Jain, Phys. Rev. B {\bf 40}, 8079 (1989).

\bibitem{wen}
X.-G. Wen, Phys. Rev. B {\bf 60}, 8827 (1999); Phys. Rev. Lett. {\bf
66}, 802 (1991)

\bibitem{barkeshli}
M. Barkeshli {\it et al. }, http://arXiv.org/abs/0910.2483.

\bibitem{dirac}
P. Dirac, Proc. R. Soc. London, Ser. {\bf A 133}, 60, (1931).

\bibitem{alvarez}
O. Alvarez, Comm. Math. Phys. {\bf 100}, 279, (1985).

\bibitem{am}
M. Alvarez and D. I. Olive, Comm. Math. Phys. {\bf 210}, 13, (2000);
Comm. Math. Phys. {\bf 217}, 331, (2001).

\bibitem{park1}
K.-S. Park and H. Han, in Preparation.

\bibitem{swingle}
B. Swingle {\it et al. }, http://arXiv.org/abs/1005.1076v1.

\bibitem{jackiw}
R. Jackiw and P. Rossi, Nucl. Phys. B {\bf 190}, 681 (1981).

\bibitem{park} K.-S. Park, {\it et al. }, Int. J. Mod. Phys. B,{\bf
23}, 4801 (2009).


\bibitem{witten2}
E. Witten, Phys. Lett. B {\bf 86}, 283 (1979).


\end{thebibliography}

\end{document}